\documentclass[letterpaper, 10 pt, journal, twoside]{ieeetran}

\usepackage{listings}
\usepackage[utf8]{inputenc}
\usepackage{algorithm}
\usepackage{graphicx}
\usepackage{float}
\usepackage{fancyhdr}
\usepackage{tabularx}
\usepackage{amsmath}
\usepackage{verbatim}
\usepackage{amsfonts}
\usepackage{amsmath}
\usepackage{mathrsfs}
\usepackage{subfigure}
\usepackage{amsthm}
\usepackage{amssymb}
\usepackage{color}
\usepackage{enumitem}
\usepackage{multirow}
\usepackage{rotating}
\usepackage{tikz}
\usepackage{pdflscape}
\usepackage{setspace}

\newcommand{\bite}{\begin{itemize}}
\newcommand{\eite}{\end{itemize}}
\newcommand{\benu}{\begin{enumerate}}
\newcommand{\eenu}{\end{enumerate}} 
\newcommand{\beq}{\begin{equation}}
\newcommand{\eeq}{\end{equation}}
\newcommand{\beqa}{\begin{eqnarray}}
\newcommand{\eeqa}{\end{eqnarray}}
\newcommand{\beqan}{\begin{eqnarray*}}
\newcommand{\eeqan}{\end{eqnarray*}}

\newcommand{\ErdosRenyi}{Erd\H{o}s-R\'enyi }
\newcommand{\nodesign}{v}

\newcommand{\htwo}{\mathcal{H}_2}

\DeclareMathOperator{\trace}{Tr}
\DeclareMathOperator{\diag}{diag}

\newtheorem{definition}{Definition}[section]
\newtheorem{theorem}{Theorem}[section]

\newtheorem{proposition}{Proposition}
\newtheorem{corollary}{Corollary}[theorem]
\newtheorem{lemma}[theorem]{Lemma}

\graphicspath{{Figures/}}

\title{Centrality measures and the role of non-normality for network control energy reduction\thanks{A preliminary version of this paper was be presented at ECC'19, \cite{lindmark2019combining}.}}

\author{Gustav Lindmark, Claudio Altafini} \date{}

\author{Gustav Lindmark and Claudio Altafini 
\thanks{G. Lindmark and C. Altafini are with the Division of Automatic Control, Dept. of Electrical Engineering, Link\"oping University, SE-58183, Link\"oping, Sweden. email: {\tt\small gustav.lindmark@liu.se, claudio.altafini@liu.se}}}%

\begin{document}

\maketitle
\thispagestyle{empty}
\pagestyle{empty}

\begin{abstract}
Combinations of Gramian-based centrality measures are used for driver node selection in complex networks in order to simultaneously take into account conflicting control energy requirements, like minimizing the average energy needed to steer the state in any direction and the energy needed for the worst direction. 
The selection strategies that we propose are based on a characterization of the network non-normality, a concept we show is related to the idea of balanced realization.
\end{abstract}

\section{Introduction}
In recent years, there has been a renewed interest in the controllability problem, motivated by its application in the context of complex networks. 
Depending on the context, many are the possible ways to define control inputs on networks, from drugs in biological networks \cite{Torres2016Drug} to dams in irrigation networks, from traffic lights in intersections  to opinion makers in social networks, etc. 
Given a network, deciding where to place the controls is often an integral part of the controllability problem. 
In the ideal case, a control can be placed on any node of the network, hence it is of interest to provide criteria for {\em driver node placement} that guarantee controllability.

The notion of structural controllability \cite{lin1974structural} has proven to be very useful to determine where to place a minimal number of driver nodes that achieve controllability (\cite{Olshevsky2014Minimal,Commault20133322} and others).
However, a network may be controllable in theory but not in practice if for instance unreasonable amounts of control energy are required to steer it in some direction.
For linear dynamics, the measures of control energy are normally formulated in terms of the controllability Gramian \cite{Muller1972Analysis}.
Several of the papers that have appeared in recent years on the subject 
in fact rely on properties of the Gramian.
For instance \cite{pasqualetti2014controllability,bof2017role} quantify the importance of the different nodes for controllability using Gramian-based network centrality measures. Optimization-based approaches are instead used in \cite{summers2016submodularity,tzoumas2016minimal}. 
None of these approaches has proven valid in all situations, especially because different measures of control energy correspond to different centrality measures and hence to different driver node selections criteria.

In \cite{Lindmark2018Minimal}, we showed numerically that the energy required to control a network is influenced by a connectivity property expressed as a ratio between the weighted outdegree and indegree of the nodes.
In this paper, the empirical results of \cite{Lindmark2018Minimal} are put into a more solid formal basis, and interpreted in terms of the algebraic properties of the adjacency matrix of the network.
Our main result is to propose two strategies for driver node placement, based on a novel characterization of network non-normality as imbalance in the distribution of energy in the network.
We establish an equivalence between a network with normal adjacency matrix and a system with balanced realization \cite{moore1981principal}.
Our formulation allows to quantify network non-normality at a node level as combinations of two different centrality metrics.
The first measure ({\em node-to-network} centrality)
quantifies the influence that each node has on the rest of the network. It corresponds to the energy with which the node excites the network.
The second measure ({\em network-to-node} centrality) describes instead the ability to control a node indirectly from the other nodes, and corresponds to the energy that reaches the node from the other nodes.
Suggestively, this centrality is formulated in terms of the observability Gramian, and it is somewhat related to structural controllability, as it identifies the nodes that cannot be controlled indirectly and hence must be driver nodes.
		
We show that the two centralities can be expressed as special cases of the $\mathcal{H}_2$ system  norm, and can be formally related to performance bounds on some of the most commonly used control energy metrics.
These results suggest that nodes with a high node-to-network centrality (i.e., with a high network influence) and nodes with a low network-to-node centrality (i.e., nodes that are difficult to control indirectly) should be driver nodes, and the strategies for driver node placement that we propose combine the centralities in such a direction.
Practically, the strategies consist in selecting the nodes that maximize the network non-normality.
In this way we achieve good performances both in terms of the average energy that is required to steer the network and in terms of the energy required to steer it in the most difficult direction.

The rest of the paper is organized as follows: In Section \ref{sec:background}, definitions are given, results on controllability are revised and different energy-related metrics are discussed.
In Section \ref{sec:centralitymeasures}, the network centralities 
are presented and their formal relations to the control energy metrics are derived.
Section \ref{sec:nonNormBal} is about network non-normality and balanced systems, while in Section~\ref{sec:ranking} the driver node placement strategies are presented. 

A preliminary version of this paper was presented at ECC'19 \cite{lindmark2019combining}. 
This conference paper discusses the network centralities for discrete-time systems. Results such as Theorem \ref{thm:WinvLB} and, most importantly, the material of Section \ref{sec:nonNormBal} are however presented here for the first time.

\section{Background}
\label{sec:background}
\subsection{Notation}

We denote $\mathbb{R}^{n\times m}$ the set of $n\times m$ matrices with real valued entries.  The $k$-th vector of the canonical basis of $\mathbb{R}^n$ is denoted $e_k,\ k\in 1,\dots, n$. 
For the vector $z\in \mathbb{R}^n$, $\|z\| = \sqrt{z^\top z}$ is its \emph{Euclidean norm}.
Given a matrix $M \in \mathbb{R}^{n\times m}$, let $M[k] = Me_k,\ k\in 1,\dots,m,$ denote the $k$-th column of $M$ and $M_{ij} = e_i^\top Me_j,\ i\in 1,\dots,n$, $j\in 1,\dots,m,$ the element on row $i$ and column $j$.
For $M\in \mathbb{R}^{n\times n}$, $\text{diag}(M) \in \mathbb{R}^n$ is the vector of its diagonal entries.
Given two matrices $M, N \in \mathbb{R}^{n\times n}$, $[M,N] = M N - N M$ is their matrix commutator.
A matrix $A$ is said {\em normal} if $[A , A^\top]=0$, non-normal otherwise. 
Given a vector $z \in \mathbb{R}^n$, the nonincreasing rearrangement of $z$ is the vector $z^\downarrow \in \mathbb{R}^n$ whose entries are the same as those of $z$ (including multiplicities) but rearranged in nonincreasing order $z_1^\downarrow \geq \dots \geq z_n^\downarrow$.

A (directed) graph $\mathcal{G}$ is indicated by the pair of its nodes and edges, $\mathcal{V} = \{\nodesign_1,\dots,\nodesign_n\}$ and $\mathcal{E} = \{(\nodesign_i,\nodesign_j),\ i,j\in 1,\dots,n\}$, or, if it is necessary to specify the edge weights, by the adjacency matrix $A$, i.e., $\mathcal{G} = \mathcal{G}(A)$. Then the weight associated with the edge from $v_i$ to $v_j$, $(v_i,v_j)$, is $A_{ji}$. 
The node $v_i\in \mathcal{V}$ is a root if it has no incoming edge and a leaf if it has no outgoing edge.

\subsection{Controllability}

We consider the following continuous-time linear time-invariant model for the network
\begin{align}
	\dot{x}(t) &= Ax(t) + B_\mathcal{K}u(t),
	\label{eq:discreteStateUpdate}
\end{align}
where $x(t) \in \mathbb{R}^n$ is the state at time $t \geq 0$, $A \in \mathbb{R}^{n\times n}$, $B_\mathcal{K} = [e_{k_1}\ \dots \ e_{k_m} ]\in \mathbb{R}^{n\times m}$ and $u(t) \in \mathbb{R}^m$. 
We represent the network with the directed graph $\mathcal{G}(A) = (\mathcal{V},\mathcal{E})$. 
Each control input is assumed to act on only one node which is then called a \emph{driver node}. The set of driver nodes is $\mathcal{K} = \{\nodesign_{k_1},\dots,\nodesign_{k_m}\} \subseteq \mathcal{V}$.
The system \eqref{eq:discreteStateUpdate} is controllable if and only if the controllability Gramian
\begin{align}
	W(t_f) &= \int_{0}^{t_f} e^{At} B_\mathcal{K} B_\mathcal{K}^\top e^{A^\top t} dt, \label{eq:dicreteTimeGramianFiniteTimeHorizon}
\end{align}
is positive definite. 
For $A$ stable, the controllability Gramian converges as $t_f \to \infty$. We omit the dependency on $t_f$ in the following.
The minimal energy that is needed to steer the network in a specific direction of the state space can be exactly computed from the controllability Gramian. 
When all directions are considered, the following metrics for the control energy are commonly used:
\begin{itemize}
	\item[\emph{i)}] The minimal eigenvalue of $W$, $\lambda_{\min}(W)$:
	The energy required to steer the system in the worst case direction is $1/\lambda_{\min}(W)$.
	\item[\emph{ii)}] $\trace(W^{-1})$: The trace of the inverse Gramian is proportional to the average energy required to control a system over all directions of the state space.
	\item[\emph{iii)}] $\trace(W)$: 
	The trace of the Gramian is inversely related to the average energy required to control a system.
\end{itemize}

See e.g. \cite{Muller1972Analysis,summers2016submodularity} for more details about the different control energy metrics.

For a stable linear input-output system $H$ with system matrices $(A, B, C)$, the $\mathcal{H}_2$ norm can be computed from the (infinite horizon) controllability Gramian,
\beq
\| H\|_2^2 =  \trace( C W C^T) .
\label{eq:H22}
\eeq


\section{Centrality measures for the control energy}
\label{sec:centralitymeasures}
We begin this section by defining a quantity we call \emph{walk energy}, which we use to derive the proposed centrality measures. 
Following that, we relate them to the considered control energy metrics.

\subsection{Centrality measures}
We define the \emph{walk energy} from $v_i$ to $v_j$ as
\begin{align}
&\varepsilon_{i\to j} = \int_{0}^{t_f} \left((e^{At})_{ji}\right)^2 dt. \label{eq:walkEnergyDef}
\end{align}
This is in fact the squared $\mathcal{H}_2$ norm of the system $A$,\ $B=e_i$,\ $C = e_j^\top$, and can be thought of as the excitation energy of node $\nodesign_j$ when a unit impulse is applied to node $\nodesign_i$.
{
	Let
	\begin{align}
		W^{(i)} = \int_{0}^{t_f} e^{At}  e_i e_i^\top e^{A^\top t} dt,\ i=1,\dots, n,
		\label{eq:gramianOfDriver}
	\end{align}
	i.e. the Gramian when $\nodesign_i$ is the only driver node. For the diagonal elements in \eqref{eq:gramianOfDriver} we have  $\left(W^{(i)}\right)_{jj} = \varepsilon_{i \to j}$, $j=1,\dots,n$.
	With the driver nodes $\mathcal{K} = \{v_{k_1},\dots,v_{k_m}\}$ the controllability Gramian \eqref{eq:dicreteTimeGramianFiniteTimeHorizon} can be written 
	\begin{align}
		W= \sum_{i=k_1,\dots,k_m} W^{(i)},
		\label{eq:WsumWi}
	\end{align}
	see e.g. \cite{pasqualetti2014controllability} for a derivation. In particular, the diagonal elements
	\begin{align}
		W_{jj} = \sum_{  i = k_1,\dots, k_m } \varepsilon_{i \to j}, \ j=1,\dots,n. \label{eq:ctrlGramDiagonalElements}
	\end{align}
}
For $A$ stable the walk energies converge as $t_f \to \infty$.

\begin{definition}
	The {\em node-to-network} centrality $p_i $ is the total walk energy from $v_i$ to all nodes,
	\begin{align}
		p_i &= \sum_{j=1}^n \varepsilon_{i\to j} = \trace(W^{(i)}).
		\label{eq:pidef}
	\end{align}
\label{def:pi}
\end{definition}
Equation \eqref{eq:pidef} is the same as the squared $\htwo$ norm \eqref{eq:H22} with $ C=I $ and $W = W^{(i)}$, hence we can interpret it as the energy injected into the system ( $ C = I$ means all nodes) by $v_i$.
We use the centrality $p_i$ for quantifying the network impact of $v_i$ as a driver node. Equation \eqref{eq:WsumWi} and the linearity of the trace operator gives 
\begin{align}
\trace(W) = \sum_{i=k_1,\dots,k_m} p_i.
\label{eq:propTrace}
\end{align}

The centrality $p= \{  p_1,  \ldots, p_n \} $ appears also in \cite{summers2016submodularity} where the driver node placement problem is investigated using optimization techniques. 
From \eqref{eq:propTrace}, for $m$ a given number of driver nodes, the control energy metric $\trace(W)$ is maximized when $\mathcal{K}$ is the set of the $m$ nodes with highest $p_i$. 
However, driver node placement based on $p$ alone does not even guarantee controllability, as worst-case directions requiring infinite energy may still exist.
For instance, the $p$ centrality does not favour roots over other nodes, although controllability is never achieved unless all roots are driver nodes. 

Introduce the fictitious output equation $y(t) = Cx(t)$, where $y(t) \in \mathbb{R}^d$ is the output at time $t$ and $C \in \mathbb{R}^{d\times n}$. The observability Gramian 
\begin{align*}
M(t_f) &= \int_{0}^{t_f} e^{A^\top t} C^\top C e^{At} dt 
\end{align*}
is positive semidefinite and converges as $t_f\to \infty$ for $A$ stable.
The {dependency on} $t_f$ is omitted in the following. 
In analogy with $W^{(i)},\ i=1,\dots n$, introduce
\begin{align*}
M^{(j)} = \int_{0}^{t_f} e^{A^\top t} e_j e_j^\top e^{A t} dt,\ j=1,\dots,n,
\end{align*}
i.e. the observability Gramian with the state of $v_j$ as the only output.
The diagonal elements are $M^{(j)}_{ii} = \varepsilon_{i\to j}$, $v_i,v_j \in \mathcal{V}$.

We use the sum of the walk energies to $v_j$ from all the other nodes as a metric for the ability to control $v_j$ indirectly,
\begin{align}
\tilde{q}_{j} = \sum_{\forall i\neq j} \varepsilon_{i\to j}. \label{eq:qtildej}
\end{align}
From the definition of walk energy we obtain $\tilde{q}_j\geq 0,\ j\in 1,\dots,n$.
The metric attains its least value $\tilde{q}_j = 0$ if and only if $\nodesign_j$ is a root. Furthermore, it is close to its minimum for nodes with only few and weak incoming edges, i.e. ``almost'' root nodes.
Besides $\tilde{q}_j$, we will also use the following centrality metric.
\begin{definition}
	The {\em network-to-node} centrality $ q_i $ is the total walk energy from all nodes to $v_j$,
	\begin{align}
		q_j &= \sum_{i=1}^n \varepsilon_{i\to j} = \trace (M^{(j)}).
	\end{align}
	\label{def:qj}
\end{definition}
The centrality $q_j$ is the squared $\htwo$ norm of the system $(A,B=I,C=e_j^\top)$, hence interpretable as the system energy that a impulse input at each node injects into node $v_j$. 
Since $q_j = \tilde{q}_j + \varepsilon_{j \to j}$ with $\varepsilon_{j \to j} > 0$, it is $q_j > 0$.

\subsection{Control energy bounds}

Lemma \ref{lem:WdiagAndq} below follows directly from \eqref{eq:ctrlGramDiagonalElements} and the definitions of $q$ and $\tilde{q}$. The result is later used to derive theoretical bounds relating the control energy metrics $\lambda_{\min}(W)$ and $\trace(W^{-1})$ to the centrality measures (Theorems \ref{thm:lambdaMinAnyK} and \ref{thm:WinvLB} respectively).

\begin{lemma}
	The diagonal elements $W_{jj},\ j=1,\dots,n,$ are bounded by
	\begin{enumerate}[label=(\roman*)]
		\item $W_{jj} = q_j$ if $\mathcal{K} = \mathcal{V}$ (the network is fully actuated), \label{qiWdiag:prop1}

		\item $0 \leq W_{jj} \leq \tilde{q}_j$ if $v_j \in \mathcal{V}\setminus \mathcal{K}$, and
		\label{qiWdiag:prop3}
		
		\item $\varepsilon_{j \to j} \leq W_{jj} \leq q_j$ if $v_j \in \mathcal{K}$.
		\label{qiWdiag:prop2}
		
	\end{enumerate}
	\label{lem:WdiagAndq}
\end{lemma}

\begin{theorem}
	With $\mathcal{K}$ a set of driver nodes, it holds
	$\lambda_{\min}(W) \leq \min\{q_i,\tilde{q}_j\}$, 
	$i = 1,\dots, n,$ and $j \text{ s.t. } v_j \in \mathcal{V}\setminus \mathcal{K}$.
	\label{thm:lambdaMinAnyK}
\end{theorem}
\begin{proof}
	Since $W$ is symmetric,
	\begin{align*}
	\lambda_{\min}(W) &= \min_{\|x\| = 1} x^\top W x \leq  e_j^\top W e_j = W_{jj},\ j = 1,\dots,n.
	\end{align*}
	The result of the theorem is obtained when this is used with properties \textit{\ref{qiWdiag:prop3}} and \textit{\ref{qiWdiag:prop2}} of Lemma \ref{lem:WdiagAndq}. 
	Notice that $\lambda_{\min}(W) \leq  q_i,\  i = 1,\dots, n,$ holds for any $\mathcal{K} \subseteq \mathcal{V}$.
\end{proof}

In the following corollary, let the indices $j_1,\dots, j_n$ be such that $\tilde{q}_{j_1} \leq \tilde{q}_{j_2} \leq \dots \leq \tilde{q}_{j_n}$.
\begin{corollary}
	For any set $\mathcal{K}$ of $m$ driver nodes, the minimal eigenvalue of the Gramian is bounded by
	\begin{align}
		\lambda_{\min}(W) \leq \min \{q_i,\tilde{q}_{j_{m+1}}\},\  i = 1,\dots, n.
		\label{eq:lambdaMinBound}
	\end{align}
	\label{cor:lambdaMin_m_q}
\end{corollary}
\begin{proof}
	Taking $\mathcal{K}$ in Theorem \ref{thm:lambdaMinAnyK} as the set of nodes with lowest $\tilde{q}_j$ gives the bound \eqref{eq:lambdaMinBound}.	
	For any other $\mathcal{K}$ s.t. $|\mathcal{K}| = m < n$ $\exists \ d \in 1,\dots, m$ s.t. $v_{j_d}\in \mathcal{V}\setminus \mathcal{K}$ and $\lambda_{\min}(W) \leq \tilde{q}_{j_d} \leq \tilde{q}_{j_{m+1}}$.
\end{proof}
As a consequence of Corollary \ref{cor:lambdaMin_m_q}, the nodes with the lowest $\tilde{q}$ and $q$ give a direct upper bound on $\lambda_{\min}(W)$, i.e. a lower bound on the energy required to control the network in the most difficult direction.

\begin{theorem}
	For any set of driver nodes $\mathcal{K}$ such that the network is controllable it holds
	\begin{align}
		\trace(W^{-1}) \geq \sum_{  j \text{ s.t. } v_j \in \mathcal{K}  } \frac{1}{q_j} + \sum_{  j \text{ s.t. } v_j \in \mathcal{V}\setminus\mathcal{K}  } \frac{1}{\tilde{q}_j}.
		\label{eq:WinvLB}
	\end{align}
	\label{thm:WinvLB}
\end{theorem}
\begin{proof}
	Since $W$ is symmetric, by the Schur-Horn theorem for Hermitian matrices \cite{horn2012matrix}, the vector of eigenvalues $\lambda(W) = [\lambda_1(W)\ \dots\ \lambda_n(W)]^\top$ majorizes $\diag(W)$, i.e.,
	\begin{align*}
	 \sum_{  i = 1 }^k \lambda_i(W)^\downarrow \geq \sum_{  i = 1 }^k \diag_i(W)^\downarrow 
	\end{align*}
	for each $k=1,2,\dots,n$, with equality for $k=n$.
	Applying Hardy-Littlewood-P\'{o}lya's inequality on majorizing sets and convex functions \cite{marshall1979inequalities} to $\lambda(W)$ and $\diag(W)$ gives
	\begin{align*}
	\sum_{  i = 1 }^n \frac{1}{\lambda_i(W)} \geq \sum_{  i = 1 }^n \frac{1}{W_{ii}}.
	\end{align*}
	
	Controllability implies that $\mathcal{V}\setminus \mathcal{K}$ contains no root nodes (all root nodes must be driver nodes). Hence, $\tilde{q}_j >0 \ \forall j$ s.t. $v_j \in \mathcal{V}\setminus \mathcal{K}$ and the bound \eqref{eq:WinvLB} exists.
	By Lemma \ref{lem:WdiagAndq}, $1/W_{jj} \geq 1/q_j$ if $v_j \in \mathcal{K}$ and $1/W_{jj} \geq 1/\tilde{q}_j$ otherwise. Hence,
	\begin{align*}
		\trace(W^{-1}) &= \sum_{  i = 1 }^n \frac{1}{\lambda_i(W)} \geq \sum_{j=1}^n \frac{1}{W_{jj}} \geq \sum_{  j \in \mathcal{K}  } \frac{1}{q_j} + \sum_{  j  \in \mathcal{V}\setminus\mathcal{K}  } \frac{1}{\tilde{q}_j}.
	\end{align*}
\end{proof}
Since $q_j \geq \tilde{q}_j + \varepsilon_{j \to j}$, the second sum in \eqref{eq:WinvLB} is the most important. Nodes with low $\tilde{q}$ that are not driver nodes result in the lower bound \eqref{eq:WinvLB} being high. 
As a corollary of Theorem \ref{thm:WinvLB} we obtain a lower bound on $\trace (W^{-1})$ for a given number of driver nodes.
\begin{corollary}
	Assume controllability. With the number of driver nodes $|\mathcal{K}| = m$ it holds
	\begin{align}
		\trace(W^{-1}) \geq \sum_{j=1}^m \frac{1}{q^\downarrow_j} + \sum_{j=1}^{n-m} \frac{1}{\tilde{q}^\downarrow_j}.
		\label{eq:trWInvIneqArbK}
	\end{align}
	A necessary but not sufficient condition for equality in \eqref{eq:trWInvIneqArbK} is that $\mathcal{K}$ are the nodes with the lowest $\tilde{q}_j$.
	\label{cor:W_inv_m}
\end{corollary}
\begin{proof}
	In Theorem \ref{thm:WinvLB}, use the fact that
	\begin{align*}
		\sum_{  j \text{ s.t. } v_j \in \mathcal{K}  } \frac{1}{q_j}
	 \geq \sum_{j=1}^m \frac{1}{q^\downarrow_j}, \ \sum_{  j \text{ s.t. } v_j \in \mathcal{V}\setminus\mathcal{K}  } \frac{1}{\tilde{q}_j}
	 \geq \sum_{j=1}^{(n-m)} \frac{1}{\tilde{q}^\downarrow_j},
	\end{align*}
	with equality if and only if $\mathcal{K}$ are the nodes with lowest $\tilde{q}_j$.
\end{proof}

%


\section{Non-normality and balanced systems}
\label{sec:nonNormBal}
The notion of balanced realization has a central role in classical control theory and is mainly used for model reduction \cite{moore1981principal}.
Here, we show that the network non-normality can be understood as imbalances in the distribution of energy in the network realization. Moreover, as we quantify these imbalances, the centralities $p$ and $q$ naturally appear. For simplicity, in this section we only consider infinite time horizon controllability and observability Gramians.

\subsection{Characterization of network non-normality}

The following definition can be found in e.g. \cite{moore1981principal}.
\begin{definition}
	A control system $(A,B,C)$ is balanced if $W = M$.
\end{definition}
In a balanced system, the states which are difficult to reach are simultaneously difficult to observe. 

If we assume a fully actuated network where the state of each node is considered an output (i.e. $B=C=I$), then any balance/imbalance is entirely due to the weighted adjacency matrix $A$.
As a matter of fact, in this case the notion of balance can be linked to the non-normality of the adjacency matrix.
Denote by $W^{(\mathcal{V})}$, $M^{(\mathcal{V})}$ the controllability and observability Gramians corresponding to $B=C=I$.
\begin{theorem}
	A stable, fully actuated and observed network is balanced if and only if the weighted adjacency matrix $A$ is normal.
	\label{thm:nonNormBalance}	
\end{theorem}
\begin{proof} 
	$W^{(\mathcal{V})}$ and $M^{(\mathcal{V})}$ are the solutions to the Lyapunov equations
	\begin{align*}
		&A^\top W^{(\mathcal{V})} + W^{(\mathcal{V})} A + I = 0, \\
		&M^{(\mathcal{V})}A^\top + A^\top M^{(\mathcal{V})} + I = 0.
	\end{align*}
	Given that $A$ is Hurwiz stable, according to Theorem 2 of \cite{barker1974normal} it holds that $W^{(\mathcal{V})} = M^{(\mathcal{V})}$ if and only if $A$ is normal.
\end{proof}
For instance undirected networks correspond to normal weighted adjacency matrices, hence they are balanced.

It follows from Theorem \ref{thm:nonNormBalance} that the matrix
\begin{align*}
N = M^{(\mathcal{V})} - W^{(\mathcal{V})} = \int_0^\infty [e^{At}, e^{A^\top t}] dt
\end{align*}
expresses the non-normality of the network. Such quantity is not invariant to a change of basis.
In particular, it is well-known \cite{moore1981principal} that for any controllable and observable system there exists a state transformation matrix $Q$ such that 
\begin{align}
	\tilde{W} =  Q^T W Q = Q^{-1}M (Q^{-1})^\top  = \tilde{M},
	\label{eq:balancngTransformation}
\end{align}
i.e. the system is balanced in the new basis.

When balancing is used for model reduction, it is also required that the two Gramians are diagonal. 
However, a change of basis leading to diagonal Gramians will in general destroy the correspondence between the elements of the state vector and the nodes, i.e. between the $A$ matrix and the network topology.	
For irreducible $A$, only a diagonal state transformation matrix preserves the topological/algebraic correspondence, as it amounts to rescaling the states of the nodes while not mixing states at different nodes.

\subsection{Non-normality in a node}
In case $N \neq 0$, we say that $z^\top N z$ is the non-normality in direction $z \in \mathbb{R}^n, ||z|| = 1$.
In particular, we can denote
\begin{align*}
	r_{\text{diff},i} = e_i^\top N e_i = M^{(\mathcal{V})}_{ii} - W^{(\mathcal{V})}_{ii}  \in \mathbb{R}
\end{align*}
the non-normality corresponding to node $\nodesign_i \in \mathcal{V}$. If $r_{\text{diff},i} = 0$ then the node $\nodesign_i$ is ``balanced'' (in the sense that it is as difficult to control as to observe). 
	
While balancing (i.e. $\tilde{W} = \tilde{M}$) cannot in general be achieved with $Q$ diagonal,
there always exists a unique positive vector, denote it $r_{\text{quot}} \in \mathbb{R}^n$
such that $Q = \diag(r_{\text{quot}})$ achieves $\diag(\tilde{M}^{(\mathcal{V})}) = \diag(\tilde{W}^{(\mathcal{V})})$,
i.e. the diagonal part of $\tilde{N} = \tilde{M}^{(\mathcal{V})} - \tilde{W}^{(\mathcal{V})}$ is canceled.
This means that the node non-normality of $\nodesign_i$ is canceled by the rescaling $\tilde{x}_i = x_i/r_{\text{quot},i}$, $ i \in 1,\dots, n$. Also $r_{\text{quot},i}$ provides a relative measure of node non-normality (with $r_{\text{quot},i} = 1$ corresponding to $\nodesign_i$ balanced). Both $r_{\text{diff}}$ and $r_{\text{quot}}$ are related to our network centralities:
\begin{theorem}
	The node non-normalities $r_{\text{diff},i}$ and $r_{\text{quot},i}$ can be expressed as $r_{\text{diff},i} = p_i - q_i$ and $r_{\text{quot},i} = (p_i/q_i)^{1/4}$, $i\in 1,\dots,n$.
\end{theorem}
\begin{proof}
	$ $\newline
	$r_{\text{diff},i}:$
	Using the cyclic property of the trace operator, it can be shown that $M^{(\mathcal{V})}_{ii} = \trace W^{(i)}$, e.g. the centrality $p_i,\ i=1,\dots,n$.
	In the same way, $W^{(\mathcal{V})}_{ii} = q_i$. Hence, $r_{\text{diff},i} = p_i - q_i$.
	
	\noindent $r_{\text{quot},i}:$ Given the condition 
	$\tilde{M}^{(\mathcal{V})}_{ii} = \tilde{W}^{(\mathcal{V})}_{ii} \ \forall i = 1,\dots,n$.
	With $Q= \diag(r_{\text{quot}})$ we obtain
	\begin{align*}
		& \tilde{W}^{(\mathcal{V})}_{ii} = q_i r_{\text{quot},i}^2 \text{ and } \tilde{M}^{(\mathcal{V})}_{ii} = p_i /r_{\text{quot},i}^2, \text{ hence} \\
		&\tilde{M}^{(\mathcal{V})}_{ii} = \tilde{W}^{(\mathcal{V})}_{ii}  \Leftrightarrow r_{\text{quot},i}^4 = p_i/ q_i		
	\end{align*}
	with $r_{\text{quot},i} = \left(p_i/q_i\right)^{1/4}$ the only positive real root.
\end{proof}

Notice that $\sum_{i=1}^n r_{\text{diff},i} = 0$, 
meaning that if some nodes have a positive non-normality then others must have a negative non-normality.

\subsection{Non-normality in a set of nodes}

The node non-normalities $r_{\text{diff},i}$ and $r_{\text{quot},i}$ can be combined for sets of nodes. For $\mathcal{S} \subseteq \mathcal{V}$, let
\begin{align}
	r_{\text{diff},\mathcal{S}} = \sum_{i \text{ s.t. } v_i \in \mathcal{S}}  e_i^\top N e_i = \sum_{i \text{ s.t. } v_i \in \mathcal{S}} r_{\text{diff},i}
	\label{eq:totalBalance}
\end{align}
be the non-normality of the node set $\mathcal{S}$.
	
Given two sets of nodes $\mathcal{S}_1\subseteq \mathcal{V}$ and $\mathcal{S}_2\subseteq \mathcal{V}$, define the net walk energy from $\mathcal{S}_1$ to $\mathcal{S}_2$ as
\begin{align*}
\Delta \varepsilon_{\mathcal{S}_1\to \mathcal{S}_2} &= 
\varepsilon_{\mathcal{S}_1\to \mathcal{S}_2} - \varepsilon_{\mathcal{S}_2\to \mathcal{S}_1} = 
\mathop{\sum_{  i \text{ s.t. } v_i \in \mathcal{S}_1  }  }_{j  \text{ s.t. } v_j \in \mathcal{S}_2}
(\varepsilon_{i \to j} - \varepsilon_{j \to i}).
\end{align*}
%
\begin{proposition}
	For $\mathcal{S} \subseteq \mathcal{V}$, the node set non-normality $r_{\text{diff},\mathcal{S}}$ is the net walk energy from $\mathcal{S}$ to $\mathcal{V}\setminus \mathcal{S}$.
\end{proposition}
The proposition follows from straight-forward manipulations of \eqref{eq:totalBalance}.

In the next section we will use the node set non-normality for driver node placement, i.e. for determining the set $\mathcal{K}$.
Since $r_{\text{diff},\mathcal{K}}$ is a linear function of the set $\mathcal{K}$ (equation \eqref{eq:totalBalance}), for a given $|\mathcal{K}| = m$, it is maximal when $\mathcal{K}$ is the set of $m$ nodes with highest {\normalfont  $r_{\text{diff},i}$}.
When $m$ is left arbitrary, \eqref{eq:totalBalance} implies that the maximal node set non-normality is given in correspondence of $\mathcal{K} = \{v_i,\ i\  \text{s.t. } r_{\text{diff},i} \geq 0 \}$.
This case describes how to partition the nodes into the two sets $\mathcal{K}$ and $\mathcal{V}\setminus \mathcal{K}$ that achieve the maximum net walk energy from the former to the latter. 

{
}

For the generalization of $r_{\text{quot},i}$ to sets of nodes, we seek a common rescaling $\tilde{x}_i = x_i/r_{\text{quot},\mathcal{S}},\ \forall i \in \mathcal{S}$, such that
\begin{align*}
	&\prod_{i \text{ s.t. } v_i \in \mathcal{S}}  \tilde{M}^{(\mathcal{V})}_{ii} = \prod_{i \text{ s.t. } v_i \in \mathcal{S}}\tilde{W}_{ii}^{(\mathcal{V})}.
\end{align*}
This is achieved for $r_{\text{quot},\mathcal{S}}$ the geometric average of $r_{\text{quot},i}$, $i \text{ s.t. } v_i \in \mathcal{S}$. Hence, for a given $|\mathcal{K}| = m$, $r_{\text{quot},\mathcal{K}}$ is maximal when $\mathcal{K}$ is the set of $m$ nodes with highest {\normalfont  $r_{\text{quot},i}$}.


\section{Driver node placement}
\label{sec:ranking}

\begin{figure}
	\centering
	\begin{tikzpicture}
		\node (img1) at (0,0) {\includegraphics[width=.7\columnwidth]{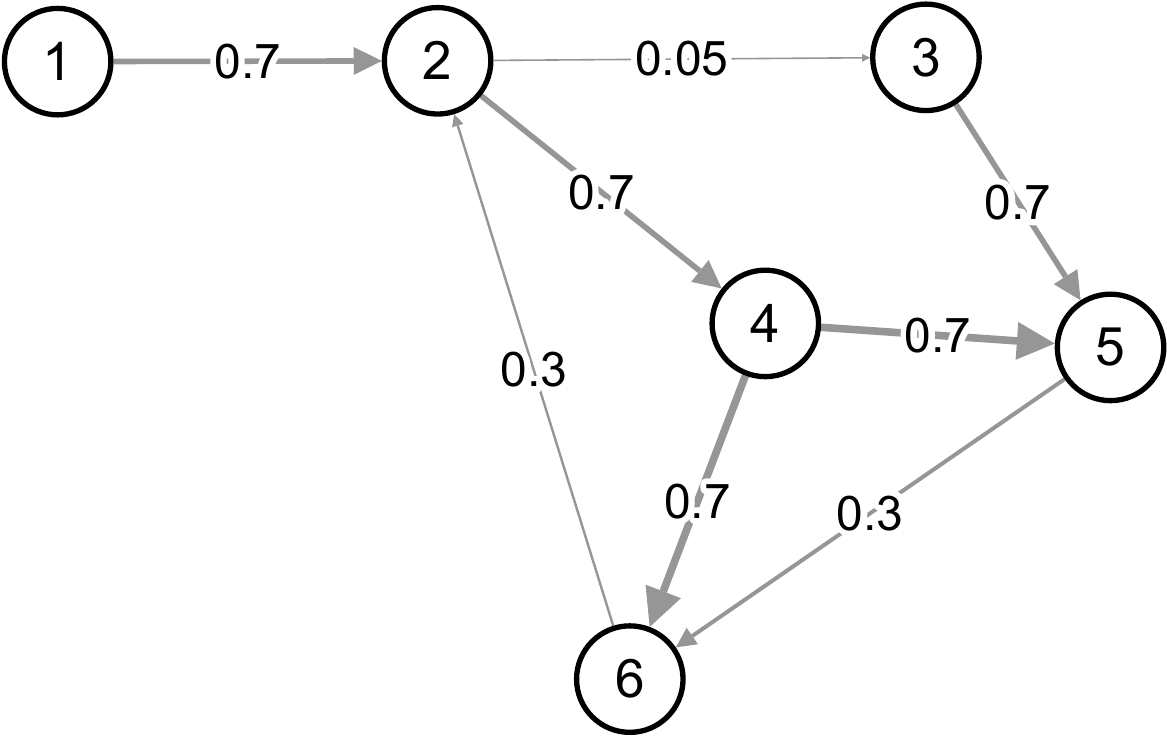}};
		\node (img2) at (-3,-.70) {
			\scriptsize
			\begin{tabular}{p{0.05\linewidth}|p{0.04\linewidth}|p{0.04\linewidth}|p{0.06\linewidth}|p{0.05\linewidth}}
			\textbf{Node}  & \boldmath$p$ & \boldmath$q$ & \boldmath$r_{\text{diff}}$ & \boldmath$r_{\text{quot}}$ \\
			\hline
			1 & 0.76          & {0.50}     & {0.26} & {1.52} \\
			2 & {0.82} & 0.71              & 0.11          & 1.15 \\
			3 & 0.64          & {0.501}    & 0.14          & {1.27} \\
			4 & {0.90} & 0.73              & {0.17} & 1.23\\
			5 & 0.53          & 0.86              & -0.33         & 0.62\\
			6 & 0.57          & 0.91              & -0.34         & 0.63\\
			\end{tabular}
		};
		\node (img3) at (-1.8,-2.8) {
			\scriptsize
			\begin{tabular}{p{0.2\linewidth}|p{0.09\linewidth}|p{0.15\linewidth}|p{0.15\linewidth}}
			\boldmath$\mathcal{K}$  & \boldmath$\trace(W)$ & \boldmath$\lambda_{\min}(W)$ & \boldmath$\trace(W^{-1})$\\
			\hline
			$\mathcal{K}_{\text{tr}} \ \ = \{2,4\}$ & 1.72 & 0  & - \\
			$\mathcal{K}_{\text{diff}} =\{1,4\}$ & 1.66 & $3.84\cdot 10^{-5}$ &  $2.80\cdot 10^4$  \\
			$\mathcal{K}_{\text{quot}} =\{1,3\}$ & 1.39 & $1.11\cdot 10^{-3}$ & $1.04\cdot 10^3$ \\
			\end{tabular}
		};
	\end{tikzpicture}
	\caption{
		A directed network with 6 nodes. All nodes have self-loops of weight $-1$ that are omitted in the figure.
		With $m = 2$ the two rankings suggest different sets of driver nodes, and the solution that maximizes $\trace(W)$ is also different.
		$\trace(W)$ is maximized with $\mathcal{K}_{\text{tr}} = \{2,4\}$. However, this choice of driver nodes renders the network uncontrollable since the root node $\nodesign_1$ is not a driver node.
		The two best nodes according to $r_{{\normalfont \text{diff}}}$ are $\mathcal{K}_{{\normalfont \text{diff}}} = \{1,4\}$ (the root node is included). They render the network controllable and $\trace(W)$ is high.
		Node 3 is ``almost'' a root, and the $r_{{\normalfont \text{quot}}}$ ranking places it as number two in importance. The resulting driver node placement $\mathcal{K}_{{\normalfont \text{quot}}} = \{1,3\}$ gives the best $\lambda_{\min}(W)$ and $\trace(W^{-1})$ but the lowest $\trace(W)$.
	}
	\label{fig:exampleNw1}
\end{figure}

We use the node non-normalities $r_{\text{diff},i}$ and $r_{\text{quot},i}$ to rank the nodes for driver node placement.
For a given $|\mathcal{K}| = m$, this means to select the set $\mathcal{K} \subseteq \mathcal{V}$ that maximizes the node set non-normality $r_{\text{diff},\mathcal{K}}$ or $r_{\text{quot},\mathcal{K}}$.
Figure \ref{fig:exampleNw1} shows a small network example with $p_i$, $q_i$, $r_{\text{diff},i}$ and $r_{\text{quot},i}$ presented for each node.

In our ranking strategies, maximization of $ p_i - q_i $ or $ p_i/q_i $ corresponds to two different trade-offs between nodes producing the largest injection of energy in the system ($ \max_i p_i$) and those relying on the least injected energy ($ \min_i q_i$).
In fact, $\max_i (p_i)$ alone corresponds to maximizing $\trace(W)$ but could correspond to ellipsoids of $ W$ which are ``squeezed'' to 0 in certain directions i.e., to infinite energy required along certain eigenspaces of $ 1/W$ (see example in Fig.~\ref{fig:exampleNw1}).
On the contrary, $ \max_i(-q_i) = \min_i (q_i)$ alone means focusing on nodes that have no or little incoming walk energy.
According to Theorems \ref{thm:lambdaMinAnyK} and \ref{thm:WinvLB} and their corollaries, these nodes should be driver nodes in order to improve $\lambda_{\min}(W)$ and $\trace(W^{-1})$. 

Observe that for a balanced network (with normal weighted adjacency matrix), $p = q$, $r_{\text{diff}}= 0$ and $r_{\text{quot}}= 1$ for all nodes, hence the rankings are degenerate.
Put differently, the best driver nodes considering the $p$ centrality are the worst nodes considering the $q$ centrality.

\renewcommand{\arraystretch}{1.2} 
\renewcommand{\arraystretch}{1}

\subsection{Simulations}
\label{sec:sim}

\begin{figure*}
	\centering
\begin{tikzpicture}
\node (img1) at (-6.3,0) {\includegraphics[width=.39\textwidth,trim=0cm 0cm 0cm 0cm]{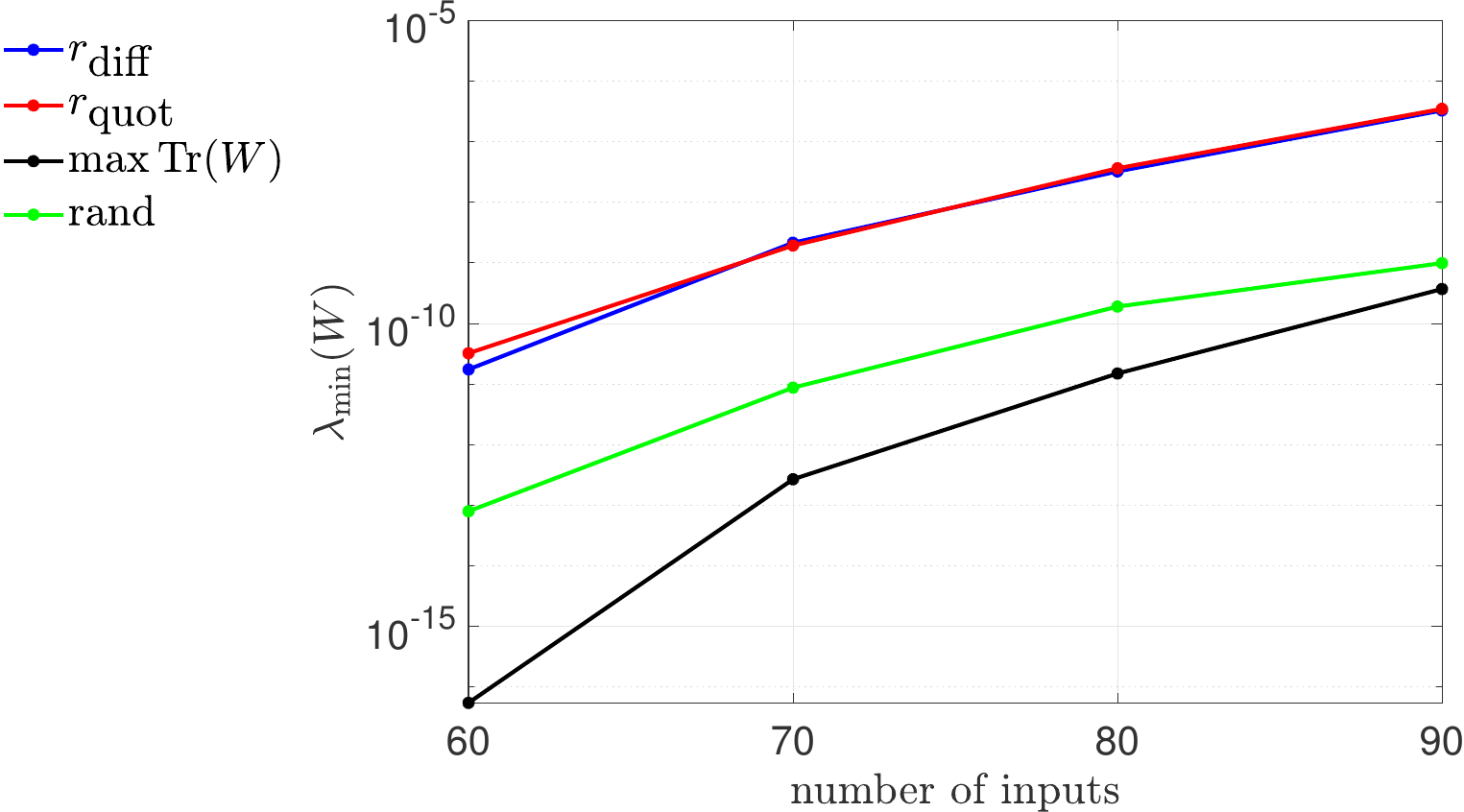}};
\node (img2) at (0,0) {\includegraphics[width=.28\textwidth,trim=0cm 0cm 0cm 0cm]{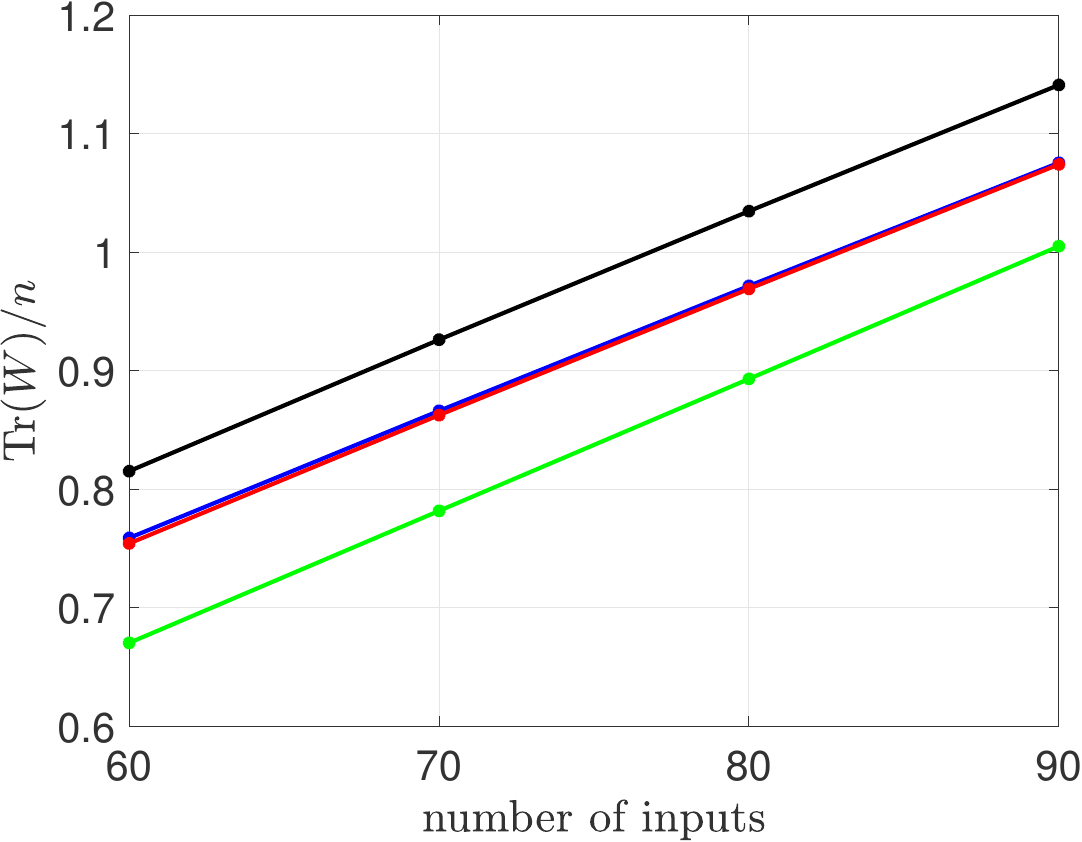}};
\node (img3) at (5.3,0) {\includegraphics[width=.28\textwidth,trim=0cm 0cm 0cm 0cm]{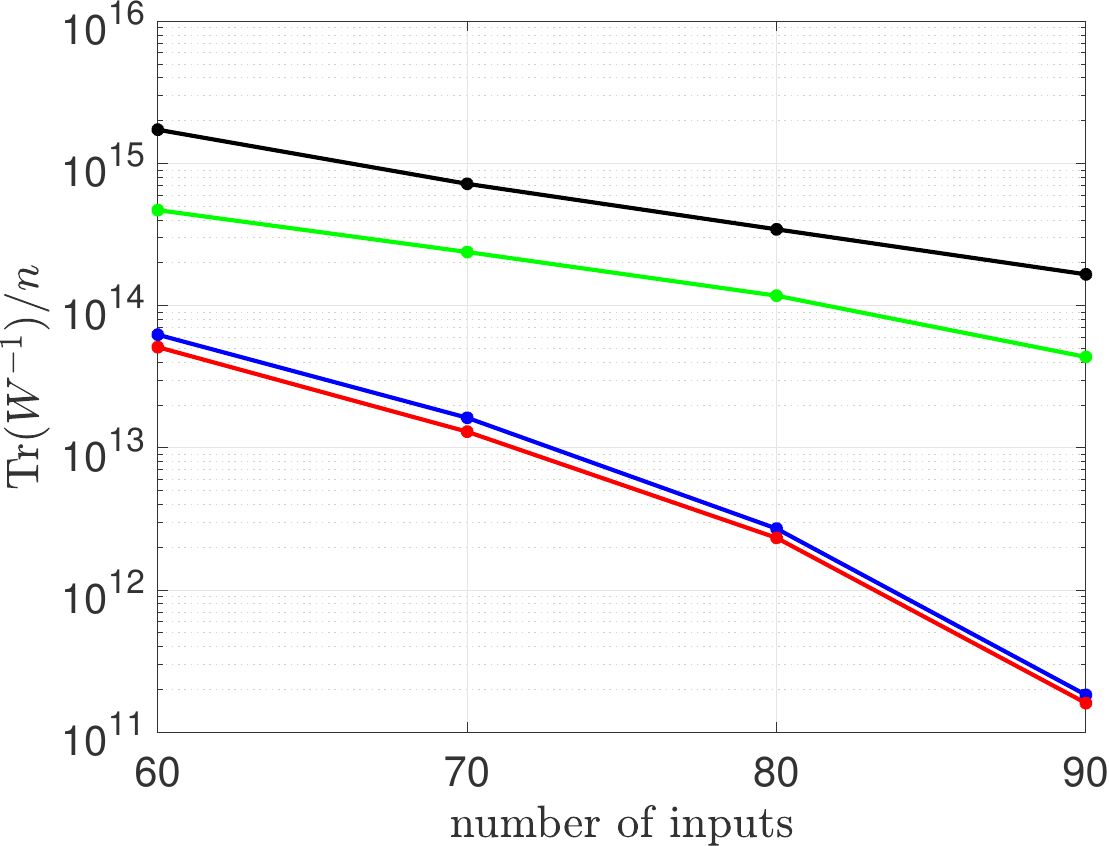}};
\node (img2) at (-9.4,0.3) {(a)};

\node (img1) at (-4.4,-4) {\includegraphics[width=.4\textwidth]{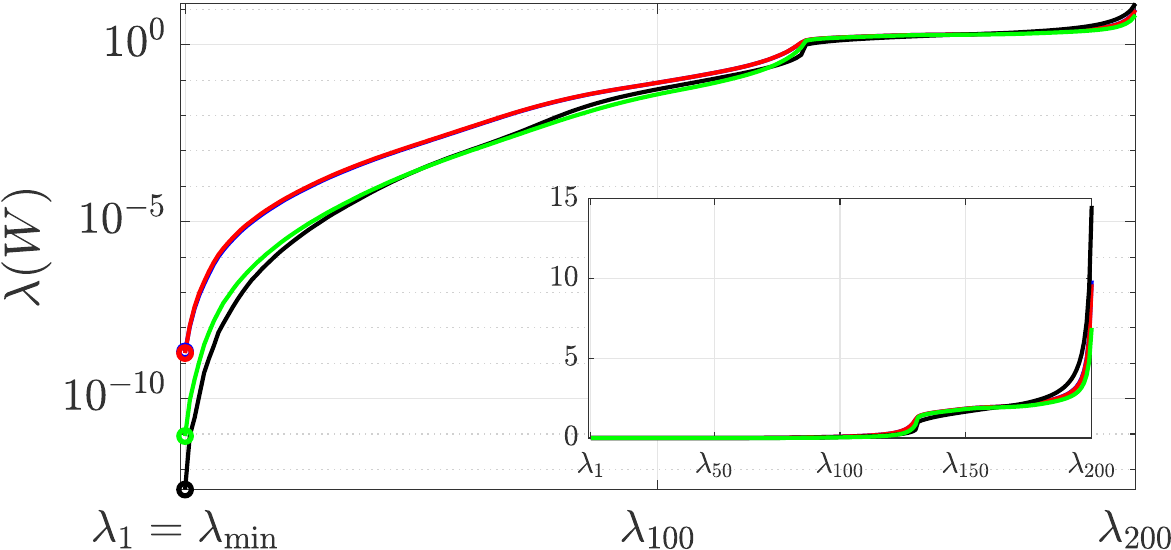}};
\node (img2) at (-8.4,-3.8) {(b)};

\node (img1) at (3.7,-4.1) {\includegraphics[width=.44\textwidth,trim=-1cm 0cm 0cm -0.3cm]{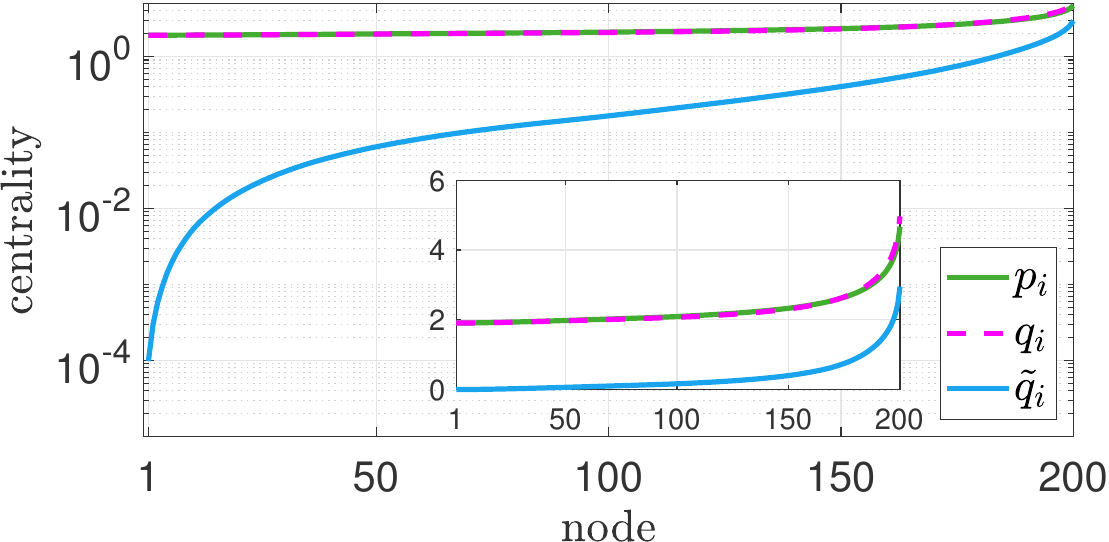}};
\node (img2) at (-0.2,-3.8) {(c)};
\end{tikzpicture}
\vspace{-.3cm}
	\caption{ 
		Simulation results for random directed scale-free networks with 200 nodes. All displayed values are the averages over 1000 network realizations. 
		(a): Control energy metrics computed for different numbers of driver nodes. Driver nodes are selected based on the proposed strategies. 
		(b): The eigenvalues of $W$ in increasing order for different ranking criteria in logarithmic and linear scale (inset). The number of driver nodes is here 70.
		(c): The metrics $p_i$, $q_i$ and $\tilde{q}_i$ in logarithmic and linear scale (inset). The values are sorted in ascending order along the x-axis.
	}
\label{fig:simulationsN500SF}
\end{figure*}

In \cite{Lindmark2018Minimal}, a variant of \normalfont{$r_{\text{quot}}$} was used for driver node placement in extensive simulation studies. 
Here, we complement these studies with an investigation 
of the amount of energy that is required to control random networks when $r_{\text{diff}}$ and $r_{\text{quot}}$ are used for driver node placement. 
For comparison, we also compute the different control energy metrics for a random driver node placement and for the placement of driver nodes that maximize $\trace(W)$. The results are presented in Figure \ref{fig:simulationsN500SF}.

We use random directed scale-free networks in our study. They have both an indegree distribution and an outdegree distribution that follows power laws.
By choosing these in a suitable way, we can obtain networks with large variations in the two network centralities $ p $ and $ q$.
The model suggested in \cite{bollobas2003directed} is used to generate random networks with 200 nodes, indegree distribution $P_{\text{in}}(k_{\text{in}}) \propto k_{\text{in}}^{-3.14}$, and outdegree distribution $P_{\text{out}}(k_{\text{out}}) \propto k_{\text{out}}^{-2.88}$. 
The edge weights are sampled from a normal distribution.
In order to ensure stability, the eigenvalues of $A$ are shifted into the complex half plane $\operatorname{Re}\{\lambda_i\} \leq -0.1, \forall i$ through the addition of negative self loops, $A:=A-\alpha I$.
As the focus is on reducing the control energy, controllability is always ensured for all choices of driver nodes by adding edges that guarantee strong connectivity when needed.

In comparison with randomly placed driver nodes, all metrics improve significantly when the driver nodes are placed according to $r_{\text{diff}}$ or $r_{\text{quot}}$; the metrics $\lambda_{\min}(W)$ and $\trace(W^{-1})$ improve several orders of magnitude.
These results are coherent with what is obtained in \cite{Lindmark2018Minimal}.
Note that the driver nodes that maximize $\trace (W)$ result in poor values of $\lambda_{\min}(W)$ and $\trace(W^{-1})$, even worse than for a random choice of driver nodes for this class of networks.
Figure \ref{fig:simulationsN500SF}(b) shows all the eigenvalues of $W$ (in increasing order) for $m=70$ driver nodes chosen according to the four criteria described above. Here $\lambda_{\min}(W)$ is the leftmost eigenvalue of each curve.
The metrics $p$, $q$ and $\tilde{q}$ are shown in Figure \ref{fig:simulationsN500SF}(c).
In fact, choosing power laws for the degree distributions means that the amount of non-normality of the corresponding adjacency matrix is large, as a significant fraction of overall outgoing edge weights is concentrated at a few nodes, and similarly for the overall incoming edge weights, thereby resulting into skewed distribution of $p_i$ and $q_i$, see Figure \ref{fig:simulationsN500SF}(c).
%
%
Corresponding results for discrete time \ErdosRenyi and directed scale free networks that are presented in \cite{lindmark2019combining} show that the improvements with $r_{\text{diff}}$ and $r_{\text{quot}}$ are smaller for \ErdosRenyi networks since their adjacency matrices have a lower degree of non-normality.

\section{Conclusions}
The network centrality measures $p$ and $q$ considered in this paper are based on system energy considerations. They reflect the fact that what makes a good driver node depends both on its influence over other nodes in the network, and on its ability to be controlled indirectly from other nodes.
These centralities are strictly related to the non-normality of the network that can be associated to the nodes. Network non-normality can be understood as imbalances in the distribution of energy in the network. For a single node it can be quantified by the difference $p-q$ or by the quotient $p/q$.
A driver node placement strategy that maximize the non-normality results in reduced energy requirements for controlling the network, i.e. all the metrics $\lambda_{\min}(W)$, $\trace(W^{-1})$ and $\trace(W)$  are simultaneously improved w.r.t. random driver node placement, although none of them is optimized. The improvements are significant for networks which have skewed in- and out-degree distributions, for which the amount of non-normality is non-negligible.


\small

\end{document}